\documentclass[twoside,english]{elsarticle}
\usepackage[LGR,LGR,T1]{fontenc}
\usepackage[latin9]{inputenc}
\usepackage{geometry}
\usepackage{textcomp}
\usepackage{amsmath}
\usepackage{amssymb}
\usepackage{setspace}
\usepackage{esint}
\makeatletter

\DeclareRobustCommand{\greektext}{%
  \fontencoding{LGR}\selectfont\def\encodingdefault{LGR}}
\DeclareRobustCommand{\textgreek}[1]{\leavevmode{\greektext #1}}
\DeclareFontEncoding{LGR}{}{}
\DeclareTextSymbol{\~}{LGR}{126}
\newcommand{\lyxmathsym}[1]{\ifmmode\begingroup\def\b@ld{bold}
  \text{\ifx\math@version\b@ld\bfseries\fi#1}\endgroup\else#1\fi}

\batchmode

\def\input@path{{\string"D:/Weberszpil/Calculo Fracional/Artigos 2013/Conexao q-calculus-Hausdorff/\string"/}}

\batchmode

\def\input@path{{\string"D:/Weberszpil/Calculo Fracional/Artigos 2013/Dirac-Pauli-G-factor/Annals of Physics/\string"/}}

\DeclareRobustCommand{\greektext}{%
  \fontencoding{LGR}\selectfont\def\encodingdefault{LGR}}
\DeclareRobustCommand{\textgreek}[1]{\leavevmode{\greektext #1}}
\DeclareFontEncoding{LGR}{}{}
\DeclareTextSymbol{\~}{LGR}{126}

\journal{}

\usepackage{babel}

\usepackage{babel}

\makeatother

\usepackage{babel}
\begin{document}

\title{\textbf{On a connection between a class of $q$-deformed algebras
and the Hausdorff derivative in a medium with fractal metric }}

\author[JW]{J. Weberszpil\corref{cor1}}

\ead{josewebe@gmail.com}

\author[MJL]{Matheus Jatkoske Lazo}

\ead{matheuslazo@furg.br}

\author[H]{J. A. Helayël-Neto}

\ead{helayel@cbpf.br}

\cortext[cor1]{Corresponding author Universidade Federal Rural do Rio de Janeiro,
UFRRJ-IM/DTL\\
 Av. Governador Roberto Silveira s/n- Nova Iguaçú, Rio de Janeiro,
Brasil, 695014.}

\address[JW]{Universidade Federal Rural do Rio de Janeiro, UFRRJ-IM/DTL\\
 v. Governador Roberto Silveira s/n- Nova Iguaçú, Rio de Janeiro,
Brasil - 695014.}

\address[MJL]{Universidade Federal do Rio Grande, FURG\\
 Instituto de Matemï¿œtica, Estatística e Física \textendash{} FURG,
Rio Grande, RS, Brazil.}

\address[H]{Centro Brasileiro de Pesquisas Físicas-CBPF-Rua Dr Xavier Sigaud
150,\\
 22290-180, Rio de Janeiro RJ Brasil. }
\begin{abstract}
Over the recent decades, diverse formalisms have emerged that are
adopted to approach complex systems. Amongst those, we may quote the
q-calculus in Tsallis' version of Non-Extensive Statistics with its
undeniable success whenever applied to a wide class of different systems;
Kaniadakis' approach, based on the compatibility between relativity
and thermodynamics; Fractional Calculus (FC), that deals with the
dynamics of anomalous transport and other natural phenomena, and also
some local versions of FC that claim to be able to study fractal and
multifractal spaces and to describe dynamics in these spaces by means
of fractional differential equations.

The question we might ask is whether or not there are common aspects
that connect these alternative approaches. In this short communication,
we discuss a possible relationship between $q-$deformed algebras
in two different contexts of Statistical Mechanics, namely, the Tsallis'
framework and the Kaniadakis' scenario, with local form of fractional-derivative
operators defined in fractal media, the so-called Hausdorff derivatives,
mapped into a continuous medium with a fractal measure. This connection
opens up new perspectives for theories that satisfactorily describe
the dynamics for the transport in media with fractal metrics, such
as porous or granular media. Possible connections with other alternative
definitions of FC are also contempled. Insights on complexity connected
to concepts like coarse-grained space-time and physics in general
are pointed out.\end{abstract}
\begin{keyword}
Hausdorff derivative \sep Fractal \sep Local fractional calculus
\sep q-deformed algebra \sep $k$-deformed algebra 
\end{keyword}
\maketitle

\section{Introduction}

The calculus with fractional derivatives and integrals of non-integer
orders started more than three centuries ago, when Leibniz proposed
a derivative of order $\frac{1}{2}$ in response to a letter from
l'Hôpital \citep{OldhamSpanier}. This subject was also considered
by several mathematicians like Euler, Fourier, Liouville, Grünwald,
Letnikov, Riemann and others up to nowadays. Although the fractional
calculus is almost as old as the usual integer order calculus, only
over the last three decades it has raised more attention due to its
applications to various fields of science (see \citep{SATM,Kilbas,Diethelm,Hilfer,Magin,SKM}
for a review). As an example, Fractional Calculus (FC) is historically
applied to study non-local or time-dependent processes, as well as
to model phenomena involving coarse-grained and fractal spaces \citep{klafter,klafter2,klafter3,9,scalas,Hilfer2,stanisvasky,Metzler,Glockle}
and frictional systems.

Presently, areas such as field theory and gravitational models demand
new conceptions and approaches which might allow us to understand
new systems and could help in extending well-known results. Interesting
problems may be related to the quantization of field theories for
which new approaches have been proposed \citep{Cresus e Everton,Kleinert,baleanu,goldfain}.

Fractional systems are described as being dissipative \citep{Gianluca - PRL 2010,Gianluca- Arxiv1106.5787}.
The use of FC is also justified here based on our proposition that
there exists an intimate relationship between dissipation, coarse-grained
media and the some limit scale of energy for the interactions. Since
we are dealing with open systems, as already commented, the particles
are in fact dressed particles or pseudo-particles that exchange energy
with other particles and the environment. Depending on the energy
scale an interaction may change the geometry of space-time, disturbing
it at the level of its topology. A system composed by particles and
the surrounding environment may be considered nonconservative due
to the possible energy exchange. This energy exchange may be the responsible
for the resulting non-integer dimension of space-time, giving rise
then to a coarse-grained medium. This is quite reasonable since, even
standard field theory, comes across a granularity in the limit of
Planck scale. So, some effective limit may also exist in such a way
that it should be necessary to consider a coarse-grained space-time
for the description of the dynamics for the system, in this scale.
Also, another perspective that may be proposed is the previous existence
of an nonstandard geometry, e.g., near a cosmological black hole or
even in the space nearby a pair creation, that imposes a coarse-grained
view to the dynamics of the open system. Here, we argue that FC allows
us to describe and emulate this kind of dynamics without explicit
many-body, dissipation or geometrical terms in the dynamical governing
equations. In some way, FC may yield an effective theory, with some
statistical average without imposing any specific nonstandard statistics.
So, FC may be the tool that could describe, in a softer way, connections
between coarse-grained medium and dissipation at a certain energy
scale.

It seems that a reasonable way to probe the classical framework of
physics is to highlight that, in the space of our real world, a generic
point is not infinitely small (or thin), it rather has a thickness.
In a coarse-grained space, a point is not infinitely thin, and here,
this feature is modeled by means of a space in which the generic differential
is not $dx$, but rather $(dx)^{\alpha}$, and likewise for the time
variable t.\textbf{ }It is noteworthy to emphasize that the notion
of fractal space-time was introduced in the 70s with the seminal work
by Stillinger \citep{Stillinger}, where the axiomatic basis for spaces
with non-integer dimension was set up. Later, the concepts associated
to a possible non-integer dimension were reinforced with the work
by Zeilinger and Svozil \citep{Zeilinger and Svozil 1985}, were they
take into account the intrinsically unavoidable finite resolution
of any physical experiment, and also the works in ref. \citep{Svozil 1987,Eyink,Palmer}
should be quoted here. Along this line, we also point out the work
by Nottale \citep{Nottale}, where the notion of fractal space-time
is revisited. Non-integer differentiability and randomness \citep{Grigolini-Rocco-West-1999}
are mutually related in their nature, in such a way that studies on
fractals on the one hand, and fractional Brownian motion on the other
hand, are often parallel as in the work of ref. \citep{Nottale}.
A function continuous everywhere, but nowhere integer-differentiable,
necessarily exhibits random-like or pseudo-random features, in that
various samplings of these functions, in the same given interval,
will be different. This may explain the huge amount of literature
extending the theory of stochastic differential equations to describe
stochastic dynamics driven by fractional Brownian motion \citep{Jumarie1,Jumarie-Lagrang Fract,Jumarie2}.
Regarding the anomalous properties of space-time with multifractal
structure, we highlight the interesting work in ref. \citep{Gianluca}
and references therein. Also, we call attention to the efforts to
build up a solid foundation for the construction of a geometry and
field theory in fractional spaces \citep{Gianluca- Arxiv1106.5787},
multifractals \citep{Gianluca- Arxiv1107.5041} and multi-scale \citep{Gianluca- Arxiv1305.3497}
space-times.

The majority of actual classical systems is nonconservative but, in
spite of that, the most advanced formalisms of classical mechanics
deal only with conservative systems \citep{RW}. Dissipation \citep{Sympletic},
for example, is present even at the microscopic level. There is dissipation
in every non-equilibrium or fluctuating process, including dissipative
tunneling \citep{Cal} and electromagnetic cavity radiation \citep{Sen},
for instance. In ref. \citep{Sympletic}, we adopt that one way to
suitably treat nonconservative systems is through FC, since it can
be shown that, for example, a friction force has its form stemming
from a Lagrangian that contains a term proportional to the fractional
derivative, which may be a derivative of any non-integer order \citep{RW}.

Parallel to the standard FC, there are the local fractional calculus
with certain definitions called local fractional derivatives, as the
ones introduced by Kolwankar and Gandal \citep{Kolwankar1,Kolwankar 2,Kolwankar 3}
with several works related to this approach; for example, the works
of refs. \citep{calculus of local fractional derivatives,On the local fractional derivative,Carpintieri},
the related approaches with Hausdorff derivative and also the formulation
with the so-called fractal derivative \citep{Hausdorff or fractal derivative,comparative-Fractal -Fractional}.
All of the mentioned approaches seem to be applicable to power-law
phenomena. There is also the recently developed $\alpha-derivative$
formalism by Kobelev \citep{Kobelev}.

Also parallel to the attempt for the description of complex systems
by FC, the $q-$calculus, in a non-extensive statistic context, has
its formal development based on the definition of deformed expressions
for the logarithm and exponential\citep{Borges 2004}, namely, the
$q-$logarithm and the $q-$exponential. In this context, an interesting
algebra emerges and the formalism of a deformed derivative opened
new possibilities for, besides the thermodynamical, other treatment
of complex systems, specially those with fractal or multifractal metrics
and presenting long-range dynamical interactions. The deformation
parameter or entropic index, $q,$ occupying an important place in
the description of those complex systems, describes deviations from
standard Lie symmetries and the formalism aimed to accommodate scale
invariance in a system with multifractal properties to the thermodynamic
formalism. For $q\rightarrow1$, the formalism reverts to the standard
one.

Here, we show that a definition of local fractional derivative by
means of mathematical limit operation is comparable to the definition
of the $q-$derivative approach and have similar rules. 

Our paper is outlined as follows: In Section 2, we present the mathematical
aspects, In Section 3, we develop the main subject of this communication
and we cast our Conclusions and Outlook in Section 4.

\section{Mathematical Aspects}

Following the works of \citep{Tsallis BJP- 20 anos,Tsallis1,Tsallis2},
once the solution to the differential equation $\frac{dy}{dx}=y,$
is the exponential function $e^{x}$, the author starts off from the
following the equation, 
\begin{equation}
\frac{dy}{dx}=y^{q},
\end{equation}
whose solution leads to the q-exponential, $y=e_{q}^{x}$. 

The $q-$derivative sets up a deformed algebra and takes into account
that the $q-$exponential is eigenfunction of $D_{(q)}$ \citep{Borges 2004}.
Borges proposed the operator for $q$-derivative as below:

\begin{equation}
{\displaystyle D_{(q)}f(x)\equiv{\displaystyle \lim_{y\to x}\frac{f(x)-f(y)}{x\ominus_{q}y}}}={\displaystyle [1+(1-q)x]\frac{df(x)}{dx}.}\label{eq:q-derivative}
\end{equation}

Here, $\ominus_{q}$ is the deformed difference operator, $x\ominus_{q}y\equiv\frac{x-y}{1+(1-q)y}\qquad(y\ne1/(q-1)).$

The differential equation, with the Hausdorff derivative proposed
in ref. \citep{Chen-Time Fabic}, is:

\begin{equation}
\frac{d^{H}y}{dx^{\alpha}}=y,
\end{equation}
and leads to the stretched exponential solution $y=e^{x^{\alpha}}$.

The fractional differential equation 
\begin{equation}
\frac{d^{\alpha}y}{dx^{\alpha}}=y,
\end{equation}
with the Caputo fractional derivative or the Modified Riemann-Liouville
approach, yields the solution in terms of the Mittag-Leffler function
$y=E_{\alpha}(x^{\alpha})$.

Now, that we have worked out these fundamental expressions, we are
ready to carry out the calculations of main interest.

\section{Fractal Continuum and the Hausdorff derivative connections}

A model that maps hydrodynamics continuum flow in a fractal coarse-grained
(fractal porous) space, which is essentially discontinuous in the
embedding Euclidean space, into a continuous flow governed by conventional
partial differential equations was suggested in ref. \citep{Tarasov}.
Using non-conventional partial differential equations based on the
model of a fractal continuous flow employing local fractional differential
operators, Balankin \citep{Balankin Rapid Comm} has suggested essentially
that the discontinuous fractal flow in a fractally permeable medium
can be mapped into the fractal continuous flow, which is describable
within a continuum framework, indicating also that the geometric framework
of fractal continuum model is the three-dimensional Euclidean space
with a fractal metric. For more details, the reader may consult refs.\citep{Balankin Rapid Comm,Balankin PRE85-Map-2012}.

Employing the local fractional differential operators in connection
with the Hausdorff derivative \citep{Chen-Time Fabic}, the latter
derivative can be written as \citep{Balankin PRE85-Map-2012}: 
\begin{eqnarray}
\frac{d^{H}}{dx^{\zeta}}f(x) & = & \lim_{x\rightarrow0}\frac{f(x\text{\textasciiacute})-f(x)}{x\text{\textasciiacute}^{\zeta}-x^{\zeta}}=\nonumber \\
 & = & \left(\frac{x}{l_{0}}+1\right)^{1-\zeta}\frac{d}{dx}f=\frac{l_{0}^{\zeta-1}}{c_{1}}\frac{d}{dx}f=\frac{d}{d^{\zeta}x}f,\label{eq:Haudorff derivative}
\end{eqnarray}
where $l_{0}$ is the lower cutoff along the Cartesian $x-$axis and
the scaling exponent, $\lyxmathsym{\textgreek{z}},$ characterizes
the density of states along the direction of the normal to the intersection
of the fractal continuum with the plane, as defined in the work \citep{Balankin PRE85-Map-2012}.

Notice that the differential equation 
\begin{equation}
\left(\frac{x}{l_{0}}+1\right)^{1-\zeta}\frac{d}{dx}y=y
\end{equation}
 leads to the stretched exponential solution $y=e^{\frac{l_{0}}{\zeta}\left(\frac{x}{l_{0}}+1\right)^{\zeta}}.$

Now, to achieve our main purposes, using the well-known gamma function,
we can remember the binomial expansion:

\begin{eqnarray}
\left(1+x\right)^{\alpha} & = & \sum_{k=0}^{\infty}\binom{\alpha}{k}x^{k}=\nonumber \\
 & = & \sum_{k=0}^{\infty}\frac{\Gamma(\alpha+1)}{\Gamma(k+1)\Gamma(\alpha-k+1)}x^{k}=\nonumber \\
 & \cong & 1+\alpha x+\frac{\alpha(\alpha-1)}{2!}x^{2}+\cdots,\label{eq:Binomial second order}
\end{eqnarray}
where $\alpha$ can be a noninteger rational number.

Developing eq.(\ref{eq:Haudorff derivative}) by the binomial series
(\ref{eq:Binomial second order}), with fractional exponent $(1-\zeta)$
to the second order, yields what follows below:

\begin{eqnarray}
\frac{d^{H}}{dx^{\zeta}}f(x) & = & \left(1+\frac{x}{l_{0}}\right)^{1-\zeta}\frac{d}{dx}f\nonumber \\
 & \cong & \left[1+\frac{(1-\zeta)}{l_{0}}x+\frac{(1-\zeta)(-\zeta)}{2l_{0}^{2}}x^{2}+\cdots\right]\frac{d}{dx}f.\label{eq:Expansion Hausdorff}
\end{eqnarray}

Comparing the first two terms of eq.(\ref{eq:Expansion Hausdorff})
with eq.(\ref{eq:q-derivative}), we can see (the nontrivial solution
for $\zeta\neq0$ ) that the entropic index, $q,$ and the scaling
exponent, $\lyxmathsym{\textgreek{z}},$ are related as folows:

\begin{equation}
[1+\frac{(1-\zeta)}{l_{0}}x]\frac{d}{dx}=[1+(1-q)x],
\end{equation}
so, 
\begin{equation}
1-q=\frac{(1-\zeta)}{l_{0}}.
\end{equation}

We can see that if $q\rightarrow1\Rightarrow\zeta\rightarrow1;$ if
$q\rightarrow0,$ than $\zeta\rightarrow1-l_{0}$ and if $l_{0}\rightarrow\infty\Rightarrow q\rightarrow1.$
So, we conclude that the deformed $q-$derivative is the first order
expansion of the Hausdorff derivative and that there is a strong connection
between these formalism by means of a fractal metric.

Now, let us examine the Kaniadakis definition of $k-$deformed derivative
\citep{Kaniadakis 1-2001}:

\begin{eqnarray}
\frac{d\, f(x)}{d\, x_{_{\{{\scriptstyle \kappa}\}}}} & = & \sqrt{1+\kappa^{2}x^{2}}\ \frac{d\, f(x)}{dx}\ \ .\label{eq:K-derivative}
\end{eqnarray}

\[
(1+\kappa^{2}x^{2})^{1/2}\cong1+\frac{1}{2!}k^{2}x^{2}-\frac{1/8}{3!}k^{4}x^{4}+\cdots.
\]
We observe that only even powers appear.

Kaniadakis $k-$deformed algebra is in a sense more complete than
the $q-$deformed algebra in the Tsallis context of non-extensive
statistics, by the fact that the former leads to a relativistic theory
and is based upon physical foundations of the principle of kinetic
interactions. In spite of that, the $k-$deformed derivative connection
with Hausdorff derivative is not so clear by a simple comparison,
but we may conjecture that there may exist some relationship between
the $k-$ and the $\zeta-$ parameters. The appearance of only even
powers in the Kaniadakis $k-$deformed derivative expansion may indicate,
whenever compared to the Hausdorff expansion, that the latter is still
more complete than the former since it contains all orders, even and
odd in the expansion. Some kind of $k-$complex number may complete
the connection with Hausdorff derivative.

Let us now examine other versions of local fractional derivative (LFD):
The controversial version of LFD in ref. \citep{Chines Yang}, that
has many similarities with the older one Jumarie'definition \citep{Jumarie1},
is

\begin{equation}
f^{(\alpha)}(x)=\left.\frac{d^{\alpha}f(x)}{dx^{\alpha}}\right|_{x_{0}}=\lim_{x\rightarrow x_{0}}\frac{\triangle^{\alpha}(f(x)-f(x_{0}))}{(x-x_{0})^{\alpha}},\label{eq:controversa}
\end{equation}
where there was taken the approximation $\triangle^{\alpha}(f(x)-f(x_{0}))\cong\Gamma(\alpha+1)\triangle(f(x)-f(x_{0})).$

Also, with the Jumarie's version of the fractional Fourier transform
as $f(x+h)=\sum_{k=0}^{\infty}\frac{h^{\alpha k}}{\Gamma(\alpha+1)}f^{(\alpha k)}(x)$,
we have for $f=x^{\alpha},$ 
\begin{eqnarray}
(x+h)^{\alpha} & = & f^{(0)}(x)+\frac{h^{\alpha}}{\Gamma(\alpha+1)}f^{(\alpha)}(x)+...\nonumber \\
 & \cong & x^{\alpha}+\frac{h^{\alpha}}{\Gamma(\alpha+1)}\Gamma(\alpha+1),
\end{eqnarray}
where we have used that the fractional derivative of a a functions
$x^{\gamma}$ is $D^{\alpha}x^{\gamma}=\frac{\Gamma(\alpha+1)x^{\gamma-\alpha}}{\Gamma(\gamma-\alpha+1)}.$

Now, with $(x+h)^{\alpha}=x'^{\alpha},$ we can write $x'^{\alpha}-x^{\alpha}\cong(x'-x)^{\alpha}.$

With these approximations, the above definition of LFD (\ref{eq:controversa})
seems to be nothing but the Balankian version of Hausdorff derivative
with $\left(\frac{x}{l_{0}}+1\right)^{1-\zeta}=\Gamma(\alpha+1)\left(\frac{x}{l_{0}}+1\right)^{1-\alpha}$.
So, we can conclude that this version is identical to the Balankin
version of Hausdorff derivative with a dilatation constant depending
on the $\alpha-$fractional parameter, and represents a mapping, similarly
to the Balankin's version.

Jumarie's \citep{Jumarie 2013,Jumarie- Acta Sin,Jumarie-Lagrang Fract,Jumarie1,Jumarie2}
definition for an alternative fractional derivative is 
\begin{eqnarray}
f^{(\alpha)}(x) & = & \left.\frac{d^{\alpha}f(x)}{dx^{\alpha}}\right|_{x_{0}}=\lim_{x\rightarrow x_{0}}\frac{\triangle^{\alpha}(f(x)-f(0))}{h{}^{\alpha}}=\\
 & = & {\displaystyle \lim_{x\rightarrow0}\, h^{-\alpha}\sum_{k=0}^{\infty}{\alpha \choose k}\,[f\left(x+(\alpha\,-\, k)h\right)-f((\alpha\,-\, k)h)].}
\end{eqnarray}

Without any prior approximation, it seems to borrow similarities from
the Gründwald-Letnikov definition and may be thought of as a weighted
sum in a linear chain of points, from $x_{0}$ to $x$, with different
weights ${\alpha \choose k}h^{-\alpha},$ for each point indexed by
$k$. An attempt to interpret this is to consider again a mapping
into a continuous medium with a fractal measure, so that the final
result for the derivative is a sum of local Hausdorff derivatives,
one for each point in a linear chain. So, the non-local character
is maintained by this weighted sum. Each local derivative in the weighted
sum is mapped into the continuous. Since in this deformed (mapped)
space the modified Leibniz product and the chain rules hold for each
local Hausdorff derivative, Jumarie's approach seems to also satisfy
those rules or, at least locally, it remains valid.

Recently, a promising new definition of LFD, called conformable fractional
derivative, has been proposed by the autors in ref. \citep{new definition}
that preserves classical properties and given by 
\begin{equation}
T_{\alpha}f(t)=\lim_{\epsilon\rightarrow0}\frac{f(t+\epsilon t^{1-\alpha})-f(t)}{\epsilon}.\label{basic}
\end{equation}

If the function is differentiable in a classical sense, the definition
above yields 
\begin{equation}
T_{\alpha}f(t)=t^{1-\alpha}\frac{df(t)}{dt}.\label{eq:Differentiable}
\end{equation}

Changing the variable $t\rightarrow1+\frac{x}{l_{0}}$, we should
write (\ref{eq:Differentiable}) as $l_{0}\left(1+\frac{x}{l_{0}}\right)^{1-\alpha}\frac{d}{dx}f$,
that is nothing but the Hausdorff derivative up to a constant and
valid for differentiable functions.

\section{Conclusions and Outtlook}

In this short communication, we have shown that the Hausdorff derivative
and the $q-$derivative, in thecontext of non-extensive statistics
share possible connections. They open up possibilities for a better
understanding of both formalisms, specially within complex systems
dynamics and with fractals and multifractals media. 

The physical basis involved in this the connection is the mapping
into the fractal continuum. Physically, this connection also justifies
the construction of some non-linear theories, where long-range interactions
between particles are present, giving a basis in the realm of complex
systems and fractals and so allowing to re-visit well-established
theories, but now with the possibility for the substitution of the
standard derivatives by the non-linear one \citep{Monteiro-Nobre Mass depend,Bruno e Borges-2014}.
This allows to introduce modifications into the equations that describe
the dynamics of such systems and gives rises to potential applications,
for example, the non-linear classical and quantum field theories,
non-linear electrodynamics and so on. Also, the entropic index that
appears in the context of non-extensive statistical theories may be
connected with the fractal dimension of the medium. The importance
of this connection is thus evident. I also indicates that local fractional
derivatives should be really relevant for studying complex systems,
interactions into the Cantor sets, porous media and son on. Going
further, we could hypothesize that the definition of the derivative,
in the appropriate context, plays an important role in the description
of complex systems, and thus, alternative definitions of local and
non-local derivative operators are relevant for the understanding
of such systems. The suiteble definition of derivatives, for example,
by the substitution of the simple translations in the independent
variable by a dilatation or power scale factor instead, may yield
more effective operators to the study of the dynamical systems with
long-range interactions. Examples of such derivatives, considering
different contexts, are the Gateaux derivative, the Fréchet derivative,
the Jackson derivative, local and non-local fractional derivatives,
the Borges-Derivative operator and so on.

We can also conjecture that these connections indicate that there
may exist some general definitions of entropy that include the Boltzmann-Gibbs
and the non-extensive statistics in a superstatiscs. 

An attempt to connect with Kaniadakis formalism has also been contemplated.

Finally, it seems to us that the Hausdorff fractal derivatives and
all the alternative versions of fractional calculus should be more
deeply investigated, under both the mathematical and physical points
of view. A better understanding of the exact differences and similarities
with respect to the traditional fractional calculus based on the Riemann-Liouville
or the Caputo definitions, and those with local fractional calculus
or even with fractional $q-$calculus \citep{Richard Herrmann,Metzler,Caceres 2004},
might be more thoroughly investigated to also determine the scope
of the applicability of them.

The authors wish to express their gratitude to FAPERJ-Rio de Janeiro
and CNPq-Brazil for the partial financial support.\textbf{\bigskip{}
}

\end{document}